# EXPLORING CITIES OF CENTRAL AND EASTERN EUROPE WITHIN TRANSNATIONAL COMPANY NETWORKS: THE CORE-PERIPHERY EFFECT

Natalia Zdanowska

Centre for Advanced Spatial Analysis, University College London,
UMR 8504 Géographie-cités, Université Paris 1 Panthéon-Sorbonne
n.zdanowska@ucl.ac.uk

**Abstract**

After the Fall of the Berlin Wall, Central Eastern European cities (CEEc) integrated the globalized world, characterized by a core-periphery structure and hierarchical interactions between cities. This article gives evidence of the 'core-periphery effect' on CEEc in 2013 in terms of differentiation of their urban functions after 1989. We investigate the position of all CEEc in transnational company networks in 2013. We examine the orientations of ownership links between firms, the spatial patterns of these networks and the specialization of firms in CEEc involved. The major contribution of this paper consists in giving proof of a core-periphery structure within Central Eastern Europe itself, but also of the diffusion of innovations theory as not only large cities, but also medium-sized and small ones are part of the multinational networks of firms. These findings provide significant insights for the targeting of specific regional policies of the European Union.

## Introduction

Cores are associated with high wages, technology, and profit inputs and outcomes. Geographically, these processes have tended to concentrate and segregate, producing places with either core or peripheral domination processes (Fujita, Thisse, 2002). 'Either defined in geographical or sociological terms, the core represents the locus of power and dominance and importantly, the source of prestige, while the periphery is subordinated to the core' (Azaryahu, 2008, p. 305).

According to the theories of polarisation of Myrdal (1957) and Hirschman (1958), the peripheral character represents an economic delay caused by the absorption effects determined by the great agglomerations. As a result, the core-periphery paradigm has become quasi-universal and started to be treated as a general law used to explain economic and political relations and interdependences of cities, based on the hypothesis that each system may be considered as having a core and a periphery (Szul, 2010). The periphery is therefore always compared to a core and the main issue is the interdependence between the two. However, the other main idea of the core-periphery model is that central nations use political and economic dominance to exploit the periphery in favour of their own interests, the relations developed being of asymmetric dependence (Szul, 2010).

In this sense, Central Eastern European cities[i] (CEEc) are generally considered as subordinated to the richer 'Western European' ones (Pascariu, Ţigănaşu, 2017). However, when analysing the orientation of CEEc economic interactions with other cities, with no doubt of less importance nowadays comparing to the exchanges of the rest of the European Union[ii], the relationships with cities from the communist past have persisted after the transition period of the 1990s and even diversified after 2000 (Zdanowska, 2018). This situation leads to question which CEEc are concerned by these different orientations of economic flows and how does the core-periphery structure of the globalized networks affect them. Are smaller cities better integrated at more regional scales compared to the larger ones capable of competing with the world-wide cities? How differently are CEEc specialized in terms of presence of firms foreignly owned? How does city-size influence the economic specialization of a city in Central and Eastern Europe?

To verify this statement this article aims at examining all cities in Central Eastern Europe (CEE) in a rarely applied network-based approach within ownership links between transnational firms at city level in 2013. Current globalization is characterized by the multiplication of flows, exchanges and networking of cities, mainly metropolises (Sassen 1991; Taylor 2004). The reduction of transportation costs, the development of new communication technologies and the dematerialization of the economy, makes transnational networks of firms' research objects to understand the relative position of cities today. Among the inter-urban networks, those built by economic links have been the subject of a limited number of studies in CEE. Many of them threat of the arrival of foreign direct investment in CEE countries (Hilber, Voicu, 2010; Pavlínek, 2004). Apart from being static, these studies use very different methods and are similar to monographs (Turnock, 2005). Many of them are of econometric nature (Karaszewski 2004; Pakulska, Poniatowska-Jaksch 2004) and employ models based on aggregated data at national level. As a result, it is difficult to make a coherent comparison between cities in CEE, and conduct a study of their interactions with other cities.

The present paper proposes to examine the size of CEEc in different of orientations of ownership links to question whether it is a determining factor for the highest FDI revenues generated. The general patterns of the networks, but also the economic specialization of the CEEc involved in these links will be analysed, while considering the role of small, medium and large cities under each of these aspects. In fact, the main contribution of this paper is to analyse all types of CEEc regardless their size. To do so the paper will be organized as follows: materials and methods, followed by the results and finally a discussion section.

**Materials and methods**

The main methods used for analysis in this article are statistical correlations, factor analysis and network indicators to understand the evolution of spatial interactions and position of CEEc. We use two different sources to explore their urban and economic functions:

- Population of cities in 2011 defined as urban agglomerations with a common and harmonized definition of cities from the *TRADEVE* database constructed for the whole Europe (Bretagnolle et al., 2016 and Guérois et al., 2019). A distinction has been made between large (more than 250,000 inhabitants), medium-sized (50,000-250,000 inhabitants) and small cities (10,000-250,000 inhabitants) in CEE [Appendix 1][iii];

- Ownership links of firms in 2013 at city level from the *ORBIS* database[iv] listing all the companies, located outside CEE, owning capital of CEE companies in all types of sectors. Additionally, information about CEE companies controlling the capital of other firms in CEE, but also outside CEE, is also included (Zdanowska, 2018). These variables permitted the reconstruction of ownership links between cities where companies are localised and to understand which CEE cities are the most concerned by these links[v].

A decomposition of the ownership links was carried out[vi] and led to the identification of capital control chains of several levels, according to the following scheme: a foreign firm (level N) controls the capital of a firm in CEE (level N-1). The latter itself owns the capital of another firm (N-2 level) [Appendix 2]. We construct an oriented network where cities (nodes) are connected by ownership ties (links) between firms, located in these cities. The orientation of the links is determined by the origin and destination of the firm in N and N-2. An aggregation of the chains, passing through the same city in N-1, was carried out according to the FDI revenues generated in this city even if the cities in N and N-2 are not necessarily the same[vii] [Appendix 3].

In order to understand, which CEEc are most present in these inter-regional networks of transnational firms, we used two indicators. The betweeness centrality computes the number of shortest paths passing through a node (Albert, Barabasi, 2002) and will allow to determine the number of times that CEEc in N-1 is a crossing point relaying capital control links towards N-2. The higher this centrality is for the same node, the greater the importance in terms of passage and role of gateway between chain levels. The degree indicator counts the number of links passing by a node, while the degree IN counts the number of incoming links (Kawa, 2013). Thus, an important degree, but a weak betweenness centrality, potentially characterizes only CEEc receiving ownership links, without relaying them thereafter.

In addition, we chose to characterize the patterns of the ownership networks regarding their more or less complex morphology in order to evaluate afterwards the size of cities implicated in these different structures. The structure of networks is defined by the set of topological, geometric and metric information (Gleyze, 2007). There are many classical measures to characterize the structure of graphs (Parlebas, 1972) – as, for example, the beta index (number of links on number of vertices) or gamma index (number of links on the number of possible links). However, they are regularly challenged because of their limitations when comparing two city graphs (Ducruet, 2010) – in our case, different capital chains. For this reason, we established our own classification of the most representative and recurrent forms of all ownership chains. This allowed us to identify six types of structures of capital control chains. We called them simple or "in chain" (Parlebas, 1972, p. 8), hierarchical in Y, polygon, star, complex hierarchical and multigroup [Appendix 4].

## Results

*Western CEEc as gateways of interactions*

The first results show that four CEE capital cities – Budapest, Prague, Bratislava and Warsaw – are the most important intermediate cities with regard to the orientations of all capital links [Figure 1]. This drives to a conclusion upon an extreme division between cities of the Western and Eastern facades of CEE in terms of their economic functions and core-periphery effect within CEE itself. Budapest is the city with the highest betwenness centrality, followed by Prague, Bratislava and Warsaw. These are playing the most important role in terms of gateways of capital links between companies. Brno, Ostrava, Cracow and Ljubljana have a lower, though consistent, betwenness centrality. On the other hand, Zagreb's, Bucharest's and Sofia's betweenness centrality is null, but their degree IN is the highest among all the cities. This means that they have only been receiving links. Połkowice, Łódź and many small towns in Slovakia are also in this situation. This division of economic functions of cities may be explained by different processes of diffusion of innovation (Hägerstrand, 1967), which affected Western capitals of CEE in the first place resulting from their geographical proximity with German, Austrian and Italian cities. It seems then that reality is much more complex than predicted by the core-periphery model as the periphery of Europe has itself very different urban dynamics.

In addition, the decomposition of capital links leads to an important observation: the privileged destination of capital control by firms in CEE at N-2 are other firms from CEE (62% of the total number of links) and not companies in the European Union (26%). Links oriented towards the post-communist space and outside Europe account for 8% and 4% respectively. In fact, the graph modeling ownership links oriented towards other CEEc (case 1, Scheme 1) is the most important in terms of size: 167 nodes and 456 links [Figure 2, (1)]. This result is contrary to what the core-periphery model would have predicted as the attraction towards the richest Western Europe – most of the time considered as the core in the model at the European scale – is not the strongest driving force of orientation of interactions. This confirms your previous observation and highlights the dynamics of phenomena occurring within the CEE region. The links are mainly concentrated and oriented towards Budapest, Bratislava, Prague and Warsaw. Their betwenness and cumulated FDI revenues are the highest among all cities[viii]. Finally, some cities with a low betweenness centrality can attract a large amount of investment income as for example, Brno, Ostrava (Czech Republic) and Ljubljana (Slovenia).

Ownership links oriented towards firms in the European Union, excluding CEECs (case 2, Scheme 1), generate four times more FDI revenue than in the previous case, which reflects still a strong 'core' effect of Western Europe on CEE, although it represents a smaller graph in terms of size (69 nodes and 205 links). The latter is more polarized on Budapest, Warsaw and Prague [Figure 2, (2)]. Prague is a passage node of the largest amount of foreign investment income (about 20 million euros), while Budapest is the gateway for the largest number of shortest paths.

The graph of links in the direction of firms in post-communist cities is much smaller in terms of the number of nodes (40) and links (60). It reveals the importance of Warsaw, but also the implication of cities from the south of CEE such as Sofia or those of the former Yugoslavia such as Sarajevo (Bosnia-Herzegovina) or Novi Sad (Serbia)[ix] [Figure 2, (3)]. This confirms that links from the communist past continue being relevant nowadays. Some CEEc (here Beliŝče, Zagreb and Sofia) are located at the intersection of links between firms in the European Union area (excluding the CEECs) and the former Yugoslavia (Serbia). We therefore, hypothesize that, in some cases, they are gateways between "Western" and "Eastern" Europe, which is a meaningful observation.

Considering the non-European orientation of the ownership links, only two chains of the capital control are part of the graph (16 nodes and 20 links) [Figure 2, (4)]. Prague (Czech Republic), Łódź (Poland), but also Świebodzin (Poland) are the most central cities[x]. The little amount of ownership links suggests that the CEEc are much less involved in strategies outside Europe, with the exception of Łódź, Świebodzin and Bucharest – absent from the other links orientations. In addition, Świebodzin is a small city, according to our classification (30,483 inhabitants in 2011), which is a type of city not appearing in other graphs' orientation.

*Small and medium-sized cities within interactions from the communist past*

The last subsection has revealed that small and medium-sized cities seem to be involved rather in smaller networks of ownership links, in terms of generated income. Their role is too often put aside by analysis on global cities (Escach, Vaudor, 2014). To verify more precisely, if the population of the CEEc is a determining factor of the orientation of interactions within transnational firms' networks, a correspondence factor analysis was carried out on a matrix of data (Dumolard, 2011), counting the number of times a CEE city in N is involved in the different orientations of links in N-2 [Figure 3, (1)].

This analysis highlights the fact that some cities mainly small and medium-sized, have links oriented only towards other CEEc – such as Banska Bystrica, Bolatice, Bralin, Breclav or Cesky Tesin [1 *, Figure 3, (1)] –, or post-communist ones – such as Belisce, Koper or Cracow [3 *, Figure 3, (1)]. On the other hand, large cities, such as Budapest, Zagreb and Warsaw, have links oriented towards European Union areas outside the CEECs, CEE and post-communist areas at the same time. This is not the case, however, for all capital-cities. Ljubljana, for example, is positioned among links oriented only towards other CEEc. Sofia is also in this case as it is present in only post-communist and CEE configurations. Bucharest, on the other hand, is characterized essentially by non-European implications – which can be explained by its geographical position close to the European Union border.

Small towns are the most representative of post-communist, Central Eastern and European Union orientation of links. Medium-sized cities are mostly present in the CEE orientations [Table 1]. This shows that large cities are not the only ones attracting capital links from abroad: small and medium-sized cities, often considered as put aside by globalization (Escach, Vaudor, 2014), are also present in these international networks.

This sub-section has highlighted the importance of including small and medium-sized cities in globalized networks – although they are not the source of the largest FDI revenues generated. This result confirms the theory of urban innovation diffusion in CEE (Hägerstrand, 1967). In fact, the largest cities in CEE have succeeded to adapt and diversify their urban and economic functions. However, smaller ones are just at a different stage of the innovation diffusion process as they are also part of the international networks.

*Size effect and network structure*

The classification of the graph structure drove to the results showing that the most recurrent structures are the simple (73 chains from 125) and the hierarchical in Y (46 chains from 125). They are polarized on a central node. As the construction of these chains implies that the node in N-1 is always a CEEc, the majority of the graphs give importance to these cities.

In fact, the star-shaped graph corresponds to the most polarized structure. The central node is an unavoidable passage for all links and its absence would lead to the disappearance of all links (Ducruet, 2010). The "hierarchical in Y" structure (Parlebas, 1972, p. 3) is the next form in terms of the importance of polarization, followed by the complex hierarchical structure. Conversely, the structure of a "multigroup" graph is a decentralized structure with several central nodes.

To analyse how cities in N-1 are involved in these different structures, we have undertaken a correspondence factor analysis. The latter is based on a matrix of data including all the CEEc in rows, the six types of structures in columns and the number of times N-1 cities are involved in these structures in values. The first two factorial axes account for 74,8% of the total inertia, which is sufficient for further interpretations. The proximity of a city to a type of structure indicates its main implication in this given structure [Figure 3, (2)].

Among the CEE capitals, Prague, Zagreb and Warsaw are rather present in chains of complex or multi-group hierarchical type. Budapest is part of complex hierarchical chains and in the only star structure, Bucharest in polygonal structures, and Bratislava in complex and hierarchical forms. Ljubljana and Sofia are involved in essentially simple type of structures as the other remaining cities. In 97% of cases, these are small or medium towns (69% and 28% respectively) [Table 2]. Conversely, large cities are the only ones to be all in polygonal or star-shaped structures. They are also the main type of cities present in complex and multi-group chains. The size effect seems to be then relevant regarding the structure of networks. This results fit in the innovation diffusion process, which conducted larger cities to be involved in more complex structures of economic networks compared to the smaller ones.

*CEEc' economic activity and city-size*

Following the same intuition as in the last subsection, we tried to understand if the diversity of economic activity is related to city-size in CEE. In this purpose, to determine the main economic specialisation of CEEc concerned by foreign capital control links, a correspondence factor analysis has been conducted based on a matrix of CEE cities, in rows, and the most respective economic specialisations of the companies in columns. An aggregation into the 9 most representative foreign capital control sectors of firms in CEE has been applied [Figure 3, (3)][xi]. The most important observation is the existence of certain cities, which are the headquarters of transnational companies specialised in high-technology intensive sectors, such as Prague, while others, such as Sofia, are rather specialized in low technology. Other cities, mainly small as Bolatice, Cherven Bryag, Decin or Galanta (except Sofia), are specialized only in single type of industries, while Warsaw or Budapest are characterized by companies from several sectors.

In fact, cities where automotive firms are located are mainly large ones (72% of cases), the media sector (75%) and real estate (100%) [Table 3]. Firms in the IT sector are exclusively located in large and medium-sized cities, with an identical distribution. In contrast, firms in the industrial sector are, in more than half of the cases, small urban centers (53%), and in 20% of the cases, medium-sized towns. The same structure is characterizing the sales sector (50% small, 21% average). A clear size effect can therefore be identified on the scale of the global value chain, ranging from the most innovative service sectors in the case of large cities to the low value-added industries, for small towns.

In addition, large cities are mainly pluri-sectoral (73%), as are some medium-sized cities (27%), while small cities are only mono-sectoral – and in 65% of cases in our sample [Table 3]. The mono-specialization of cities, in a type of activity, results from a new cycle of innovations having "specifically selected a particular group of cities by specializing them in relation to the rest of the urban system" (Bretagnolle, Pumain, 2010, p. 8). With regard to the CEEc, during the industrial revolution it has concerned small towns in the old mining and steel basins of Silesia, but also in the big industrial cities created during the communist period. The multi-sectoral cities are mainly capital cities. The latter are also the ones that gave birth to new technologies of the nineteenth century, stimulating the development of high-level services, particularly in the automotive sector (Bohan, 2016).

**Discussion**

This paper aimed at observing the effect of integration of CEEc into globalized networks after 1989. We conclude upon an extreme division between cities of Central Eastern Europe in terms of their urban functions within ownership networks of firms in 2013, confirming a core-periphery effect in CEE itself.

A strong opposition has been revealed between cities of the Western and Eastern facade of CEE in terms of centrality measures in networks. Budapest is the city with the highest betweenness centrality, followed by Prague, Bratislava and Warsaw. These are playing the most important role in terms of gateways of capital links between companies. All other cities are mainly only receiving or are at the origin of ownership links. This observation has been confirmed within different orientations of links, although small and medium-sized cities are more present in the post-communist and Central Eastern European orientations of ownership links. This highlights the importance of including these cities in globalized networks – although they are not the source of the largest FDI revenues generated.

Further analysis showed a differentiation between small (10,000-50,000), medium (50,000-250,000 inhabitants) and large cities (more than 250,000 inhabitants), according to several aspects. In terms of ownership networks structures in which cities are involved in N-1, small cities are mainly present in simple ones, while large cities are involved in more complex networks (hierarchical, star). Moreover, in terms of the economic specialization of firms, small and medium-sized cities are mainly mono-sectorial and specialize in the low-intensive industry, while larger ones are multi-sectoral and specialized in high-intensive sectors as finance and media.

We thus spot a size effect that oscillates along the global value chain of the economy, ranging from the most innovative service sectors for large cities to the low-intensive industry for smaller cities. All these result confirms the theory of urban innovation diffusion (Hägerstrand, 1967). In fact, mainly the largest cities in CEE have succeeded to adapt and diversify their urban and economic functions. However smaller ones are not excluded from globalization processes and are just at a different stages of the innovation diffusion. Thus, all the observations in this paper seem to confirm already existing models of cities: on the one hand, "metropolises", and on the other, "specialized" cities (Cattan, Saint Julien, 1998).

The paper's results on extreme oppositions between CEEc also permit to open a discussion regarding the application of the core-periphery model in the context of CEE. The analysis has revealed that some dynamic cities as Budapest, Warsaw, Bratislava, Prague play an important role at the European scale as gateway between the "West and the East". Other smaller cities from the Eastern facade of CEE still remain mono-sectoral. This drives to the conclusion of existence of several cores and peripheries in Europe and more specifically within Central Eastern Europe itself. These findings provide significant insights for the targeting of specific regional policies of the European Union.

## References

Albert R and Barabasi AL (2002) Statistical mechanics of complex networks. *Review of Modern Physics* 74: 45–97.

Azaryahu M (2008) Tel Aviv: Center, periphery and the cultural geographies of an aspiring metropolis. *Social & Cultural Geography* 9(3): 303–318.

Bohan C (2016) *Les stratégies des firmes multinationales automobiles dans les villes de l'élargissement européen: réseaux urbains et organisation en chaîne globale de valeur* [Multinational companies' strategies in cities of European Enlargement: urban networks and organisation in the global value chain]. PhD Thesis. University of Lausanne, Switzerland.

Brandes U (2001) A Faster Algorithm for Betweenness Centrality. *Journal of Mathematical Sociology* 25(2): 163–177.

Bretagnolle A, Giraud T, Guérois M and Mathian H (2012) A new database for the cities of Europe? Urban Morphological Zones (CLC2000) confronted to three national databases of urban agglomerations (Denmark, France, Sweden). *Environment and Planning B* 39(3): 439–458.

Bretagnolle A, Guérois M, Pavard A, Gourdon P, Zdanowska N et al. (2016) Demographical Trajectories of European urban areas (1961-2011) (TRADEVE)*.* Report, University Paris 1 Panthéon-Sorbonne, France.

Bretagnolle A and Pumain D (2010) Comparer deux types de systèmes de villes par la modélisation multi-agents. In: Weisbuch G and Zwirn A (eds) *Qu'appelle-t-on aujourd'hui les sciences de la complexité?* Langages, réseaux, marchés, territoires. Paris: Vuibert, pp. 271–299.

Cattan N and Saint-Julien T (1998) Modèles d'intégration spatiale et réseaux des villes en Europe occidentale. *L'Espace géographique* 27(1): 1–10.

Ducruet C (2010) *Les mesures globales d'un réseau: Version 1*, In: Groupe f.m.r. (flux, matrices, réseaux). Available at: https://halshs.archives-ouvertes.fr/halshs-00541902.

Dumolard P (2011) Données géographiques. Analyse statistique multivariée. Paris: Éditions Lavoisier, Publications Hermès Sciences.

Escach N and Vaudor L (2014) Réseaux de villes et processus de recomposition des niveaux: le cas des villes baltiques. *Cybergeo: European Journal of Geography [En ligne], Politique, Culture, Représentations* 679.

Fujita M and Thisse J-F (2002) *Economic of agglomeration, cities, industrial location and regional growth*, Cambridge: Cambridge University Press.

Gleyze J-F (2007) Effets spatiaux et effets réseau dans l'évaluation d'indicateurs sur les nœuds d'un réseau d'infrastructure. *Cybergeo: European Journal of Geography [En ligne], Systèmes, Modélisation, Géostatistiques* 370.

Guérois M, Bretagnolle A, Pavard A, Gourdon P and Zdanowska N (2019) Following the population of European urban areas in the last half century (1961-2011): the TRADEVE database. *Cybergeo: European Journal of Geography [En ligne], Espace, Société, Territoire* 891.

Hägerstrand T (1967) *Innovation diffusion as a spatial process*. Chicago: University of Chicago Press.

Hilber CAL and Voicu I (2010) Agglomeration economies and the location of foreign direct investment: Empirical evidence from Romania. *Regional Studies* 44(3): 355–371.

Hirschman A (1958) *The Strategy of Economic Development*. New Haven: Yale University Press.

Karaszewski W (2004) *Bezpośrednie inwestycje zagraniczne. Polska na tle świata* [Foreign Direct Investment. Poland at the world-wide scale]. Toruń: Dom Organizatora.

Kawa A (2013) Analiza sieci przedsiębiorstw z wykorzystaniem metody SNA [Company networks analysis with SNA method application]. *Przedsiębiorczość i Zarządzanie* 14(13): 77–87.

Myrdal G (1957) *Economic Theory and Underdeveloped Regions*. London: Gerald Duckworth.

Pakulska T and Poniatowska-Jaksch M (2004) *Bezpośrednie inwestycje zagraniczne w Europie Środkowo-Wschodniej. Koncentracja kapitału zagranicznego w Polsce* [Foreign Direct Investments in Central Eastern Europe. Concentration of foreign capital in Poland]. Monografie i Opracowania 519. Warszawa: Szkoła Główna Handlowa.

Pascariu G and Ţigănaşu R (2017) Integration, Growth and Core-Periphery Pattern in EU's Economy: Theoretical Framework and Empirical Evidences. In Pascariu G, Pedrosa Da Silva Duarte MA (eds) *Core-Periphery Patterns Across the European Union*. Bingley: Emerald Publishing Limited, pp. 23–85.

Pavlínek P (2004) Regional development implications of foreign direct investment in Central Europe. *European Urban and Regional Studies* 11(1): 47–70.

Sassen S (1991) *The Global City: New York, London, Tokyo*. Princeton (NJ): Princeton University Press.

Śleszyński P (2007) *Gospodarcze funkcje kontrolne w przestrzeni Polski* [Economic control functions in the Polish space]. *Prace Geograficzne* 213. Warsaw: Instytut Geografii i Przestrzennego Zagospodarowania. Polska Akademia Nauk.

Szul R (2010) Światowy system polityczny. Struktury i idee [World political system. Structures and ideas]. *Studia Społeczne* 2-3: 47–76.

Taylor PJ (2004) *World City Network: A Global Urban Analysis*. Abingdon-On-Thames: Routledge.

Turnock D (2005) *Foreign Direct Investment and Regional Development in East Central Europe and the Former Soviet Union*. Ashgate: Ashgate Pub.

Zdanowska N (2018) *Intégration des villes d'Europe centrale et orientale dans l'économie-monde depuis 1989: une entrée par les réseaux internationaux de commerce, de transport aérien et de firmes* [Integration of Central Eastern European cities in the world-economy since 1989 in the light of transnational trade, air traffic and firm networks]. PhD thesis, Université Paris 1 Panthéon-Sorbonne, France.

# Figures

**Figure 1**. Degree IN and betwenness centrality of CEEc within all orientations of ownership links in N-2 [xii]

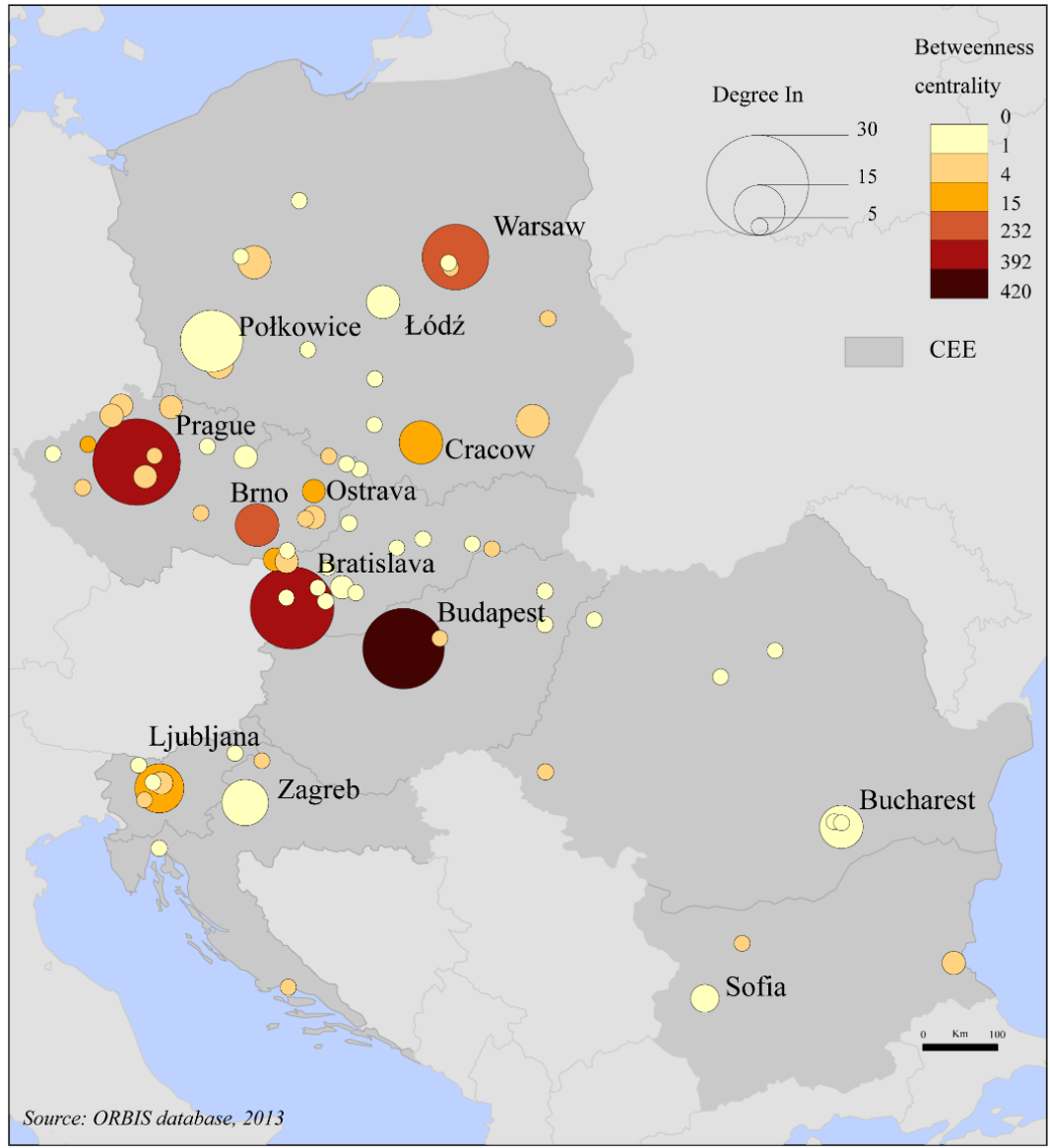

**Figure 2.** Network of ownership links oriented towards other cities of CEE, European Union, the post-communist space and outside Europe (N-2 level)[1]

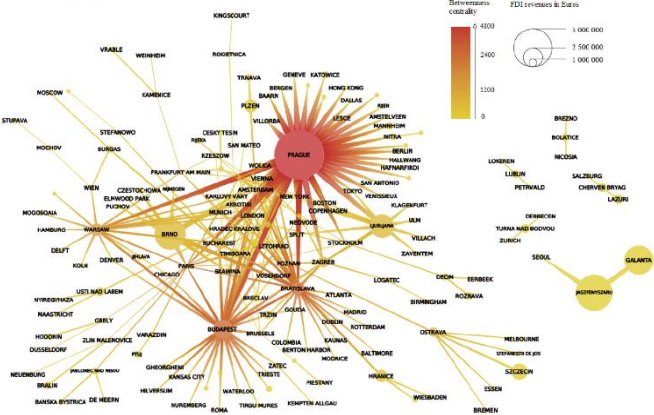
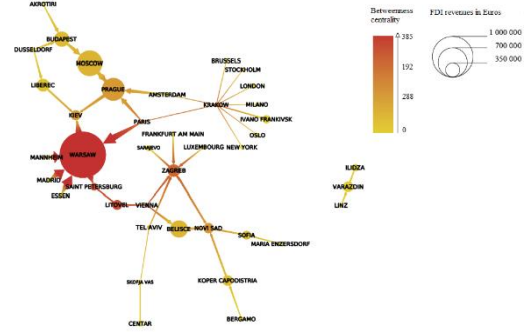
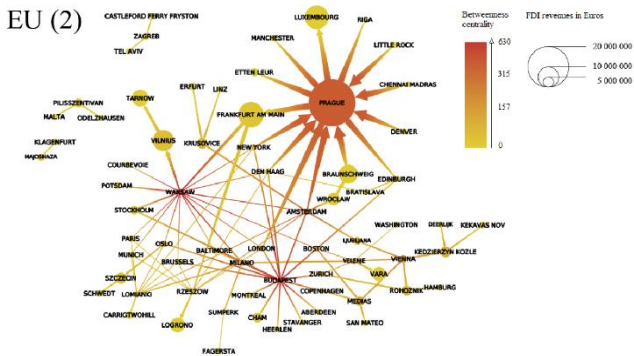
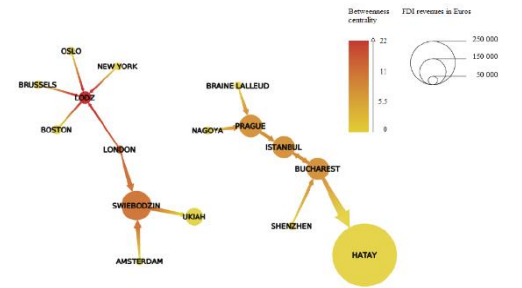

---

[1] Tulip software and the Hachul and Junger's FM3 specialization algorithm have been used for constructing these graphs. When two cities – for example Warsaw and Budapest – are related by a link in two directions, Tulip's representation will result in a double-thick link. Link arrows are visible only by zooming in on a particular node.

**Figure 3.** CEEc' implication according to: four orientations of the ownership links in N-2 (1), six structures of ownership networks (2), nine mains economic sectors of firms concerned by ownership links in N-1 (3).

(1)

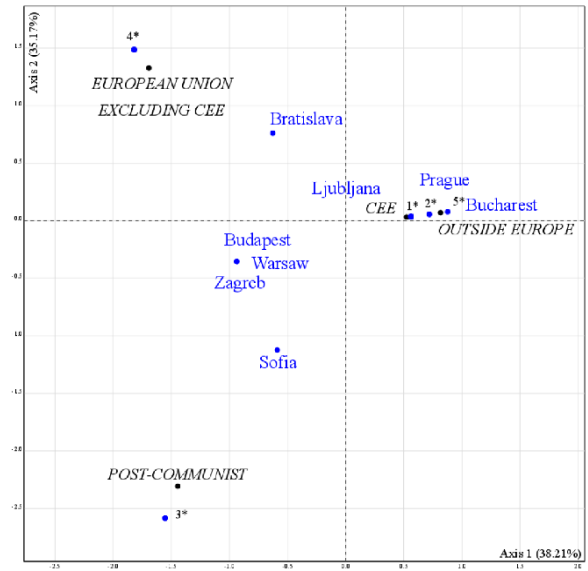

LINKS ORIENTATION IN N-2 ■  LARGE CITIES ■  MEDIUM-SIZED CITIES ■  SMALL CITIES ■

1* Banska Bystrica, Bolatice, Breclav, Brno, Bourgas, Cesky Tesin, Cherven Bryag, Częstochowa, Decin, Galanta, Gbely, Gheorgheni, Hodonin, Hradec Kralove, Hranice, Jablonec nad Nisou, Jaszfenyszaru, Jihlava, Kamenice, Karlovy Vary, Katowice, Lazuri, Legnica, Lesce, Letohrad, Logatec, Lublin, Medvode, Mochov, Modrice, Mogosoaia, Nitra, Nyiregyhaza, Ostrava, Petrvald, Piestany, Plzen, Polkowice, Poznań, Ptuj, Puchov, Rijeka, Rokietnica, Roznava, Skawina, Split, Stefanestii de Jos, Stefanovo, Stupava, Szczecin, Tirgu Mures, Trnava, Turna nad Budvou, Usti nad Labem, Vrable Wolica, Zatec, Zlin

2* Łódź

3* Belisce, Cracow, Koper, Liberec, Skofja Vas

4* Kędzierzyn Koźle, Krusovice, Majohaza, Medias, Pilisszentiva, Rohoznik, Rzeszów, Sumperk, Tarnów, Velenje, Wrocław

5* Świebodzin

(2)

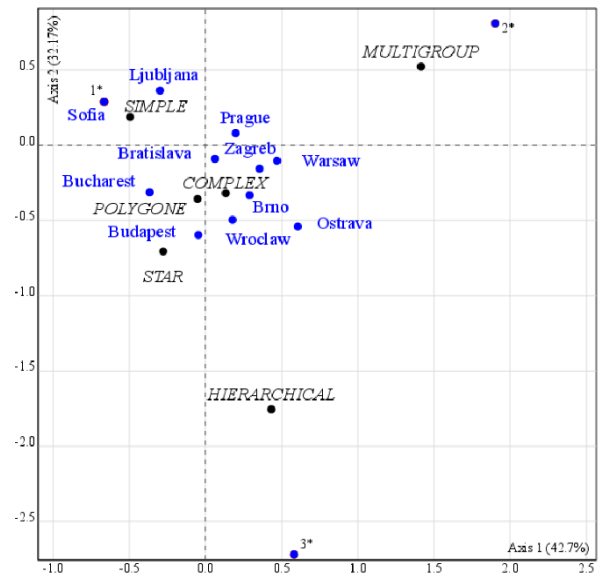

TYPE OF STRUCTURE ■  LARGE CITIES ■  MEDIUM-SIZED CITIES ■  SMALL CITIES ■

1* Banska Bystrica, Belisce, Bolanice, Breclav, Bourgas, Bydgoszcz, Centar, Cherven Bryag, Częstochowa, Galanta, Hradec Kralove, Hranice, Jabloniec nad Nisou, Jeszfenyszaru, Jihlava, Karlovy Vary, Kędzierzyn, Koźle Krusovice, Lazuri, Letohrad, Liberec, Logatec, Lublin, Kamenice, Katowice, Koper, Majoshaza, Medvedove, Mochov, Modrice, Mogosoaia, Nitra, Nyiregyhaza, Ostrokowice, Piestany, Pilisszntiva, Plzen, Ptuj, Puchov, Skofja vas, Split, Stefanovo, Stupava, Sumperk, Szczecin, Timisoara, Tirgu Mures, Trnava, Trzin, Usti nad Labem, Varazdin, Vrable, Wolica, Zatec, Zlin

2* Cesky Tesin, Gheorgheni, Cracow, Legnica, Łódź, Medias, Polkovice, Poznań, Rijeka, Rzeszów, Skawina, Świebodzin, Tarnów, Valenje

3* Decin, Gbely, Hodonin, Roznava, Rohoznik, Stefanestii de Jos, Turna nad Bodvou

(3)

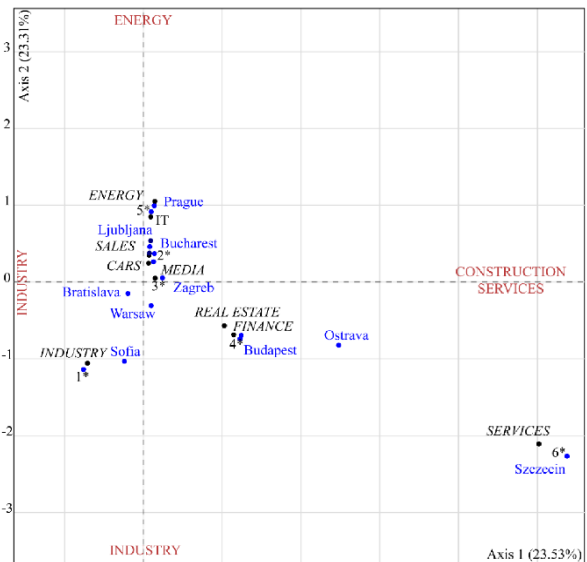

SECTORS ■  LARGE CITIES ■  MEDIUM-SIZED CITIES ■  SMALL VILLES ■

1* Bolatice, Cherven Bryag, Częstochowa, Decin, Galanta, Gheorgheni, Hranice, Jablonec nad Nisou, Jaszfenyszaru, Jihlava, Karlovy Vary, Lazuri, Lesce, Letohrad, Logatec, Lublin, Medvode, Petrvald, Plzen, Stupava, Trnava, Turna nad Bodvou, Usti nad Labem

2* Banska Bystrica, Bralin, Breclav, Brezno, Brno, Hradec Kralove, Kamenice, Katowice, Mochov, Mogosoaia, Piestany, Puchov, Roznava, Stefanestii de Jos, Trzin, Voluntari, Vrable, Zatec

3* Bydgoszcz, Legnica, Split, Timisoara

4* Legnica, Łódź, Tirgu Mures, Wolica

5* Bourgas, Rijeka, Nitra

6* Ptuj, Varazin

**Tables**

**Table 1.** Number of small, medium-sized and large cities in N-1 implicated in ownership links oriented in N-2 towards the European Union (EU), Central and Eastern Europe (CEE), the post-communist space (PC) and outside Europe (OE)

|  |  | Orientation in N-2: | | | |
|---|---|---|---|---|---|
|  |  | EU | CEE | PC | OE |
| City in N-1 | SMALL | 47 | 56 | 50 | 14 |
|  | MEDIUM-SIZED | 24 | 29 | 10 | 0 |
|  | LARGE | 29 | 15 | 40 | 86 |

**Table 2.** Small, medium and large cities in N-1 according to the different types of capital chain structures (in %)

|  | SMALL | MEDIUM-SIZED | LARGE | TOTAL |
|---|---|---|---|---|
| SIMPLE | 69 | 28 | 4 | 100 |
| HIERARCHICAL | 86 | 14 | 0 | 100 |
| POLYGONE | 0 | 0 | 100 | 100 |
| STAR | 0 | 0 | 100 | 100 |
| COMPLEXE | 0 | 29 | 71 | 100 |
| MULTIGROUP | 39 | 22 | 39 | 100 |

**Table 3.** Share of small, medium and large cities by N-2 firm sectors of ownership links (in%)

|  | SMALL | MEDIUM-SIZED | LARGE | TOTAL |
|---|---|---|---|---|
| AUTOMOTIVE | 14 | 14 | **72** | 100 |
| FINANCE | 9 | 18 | **73** | 100 |
| IT | 0 | **50** | 50 | 100 |
| INDUSTRY | **53** | 20 | 27 | 100 |
| MEDIA | 8 | 17 | **75** | 100 |
| REAL ESTATE | 0 | 0 | **100** | 100 |
| SALES | **50** | 21 | 29 | 100 |
| SERVICES | 29 | 14 | **57** | 100 |
| ENERGY | 33,3 | 33,3 | 33,3 | 100 |
| MONO-SECTORAL | **65** | 24 | 11 | 100 |
| PLURI-SECTORAL | 0 | 27 | **73** | 100 |

# Appendix

**Appendix 1.** Large, medium and small cities in CEE in 2011

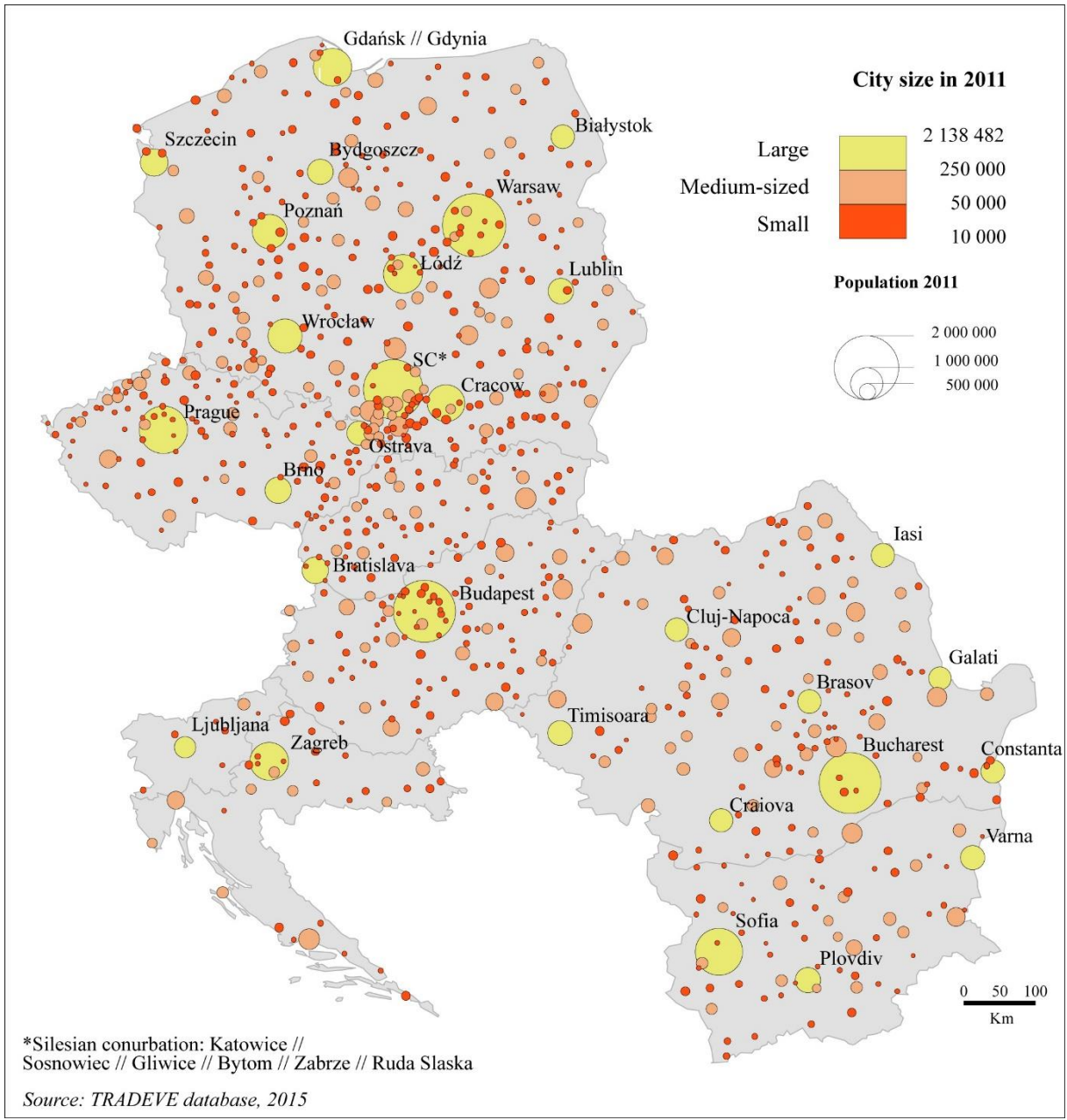

**Appendix 2.** Decomposition of the capital control links of companies in CEE into several levels

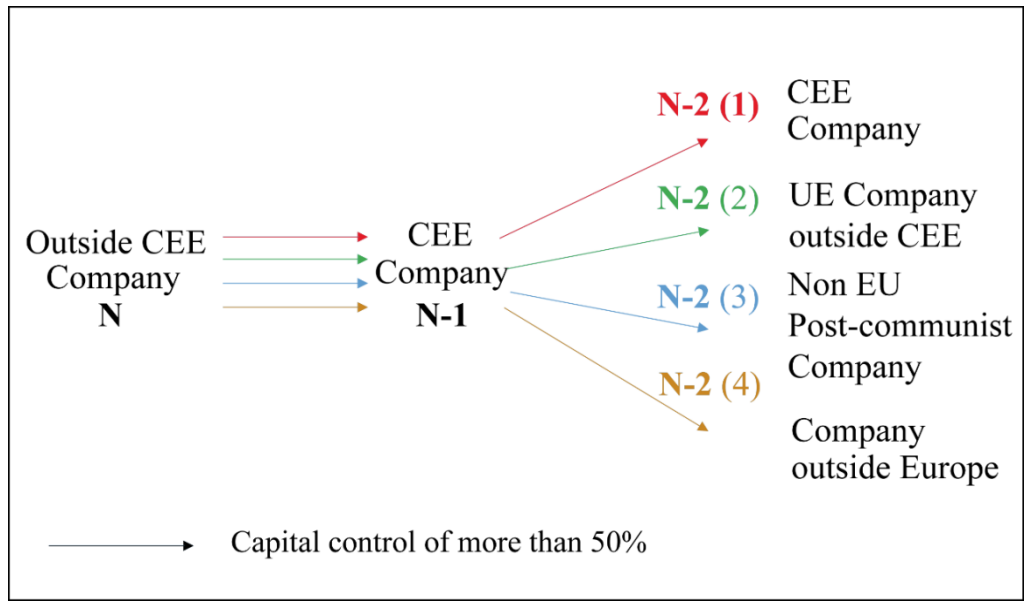

**Appendix 3.** Compilation of ownership chains between firms at city level

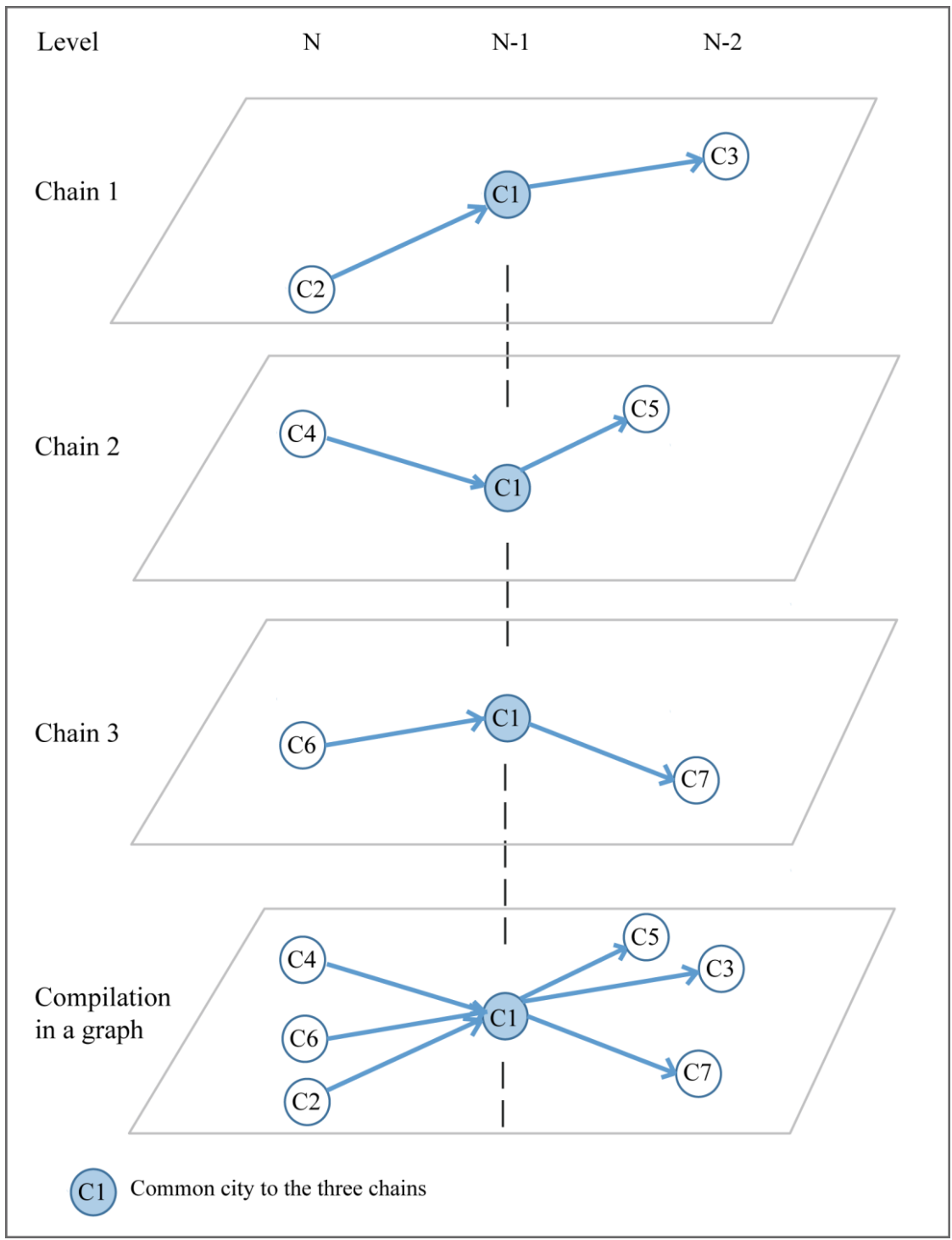

**Appendix 4.** Forms and number of capital chains after decomposing ownership links into 125 chains

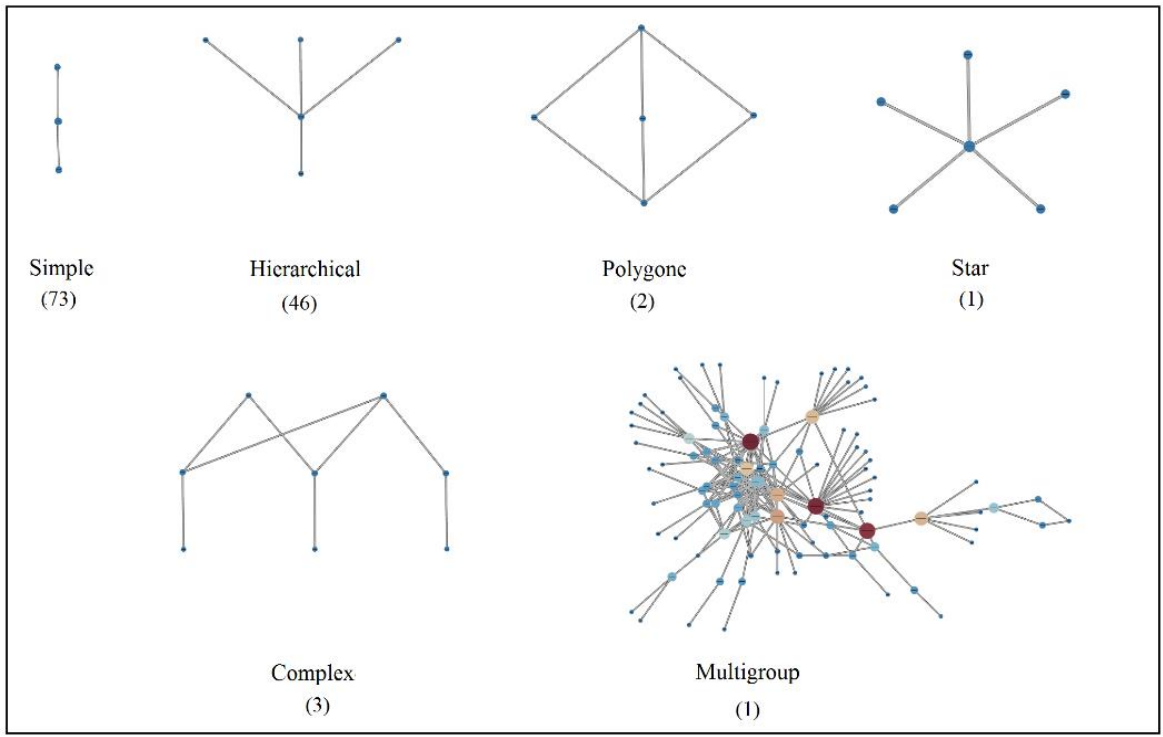

Simple (73)　　Hierarchical (46)　　Polygone (2)　　Star (1)

Complex (3)　　Multigroup (1)

i   Defined as cities of post-communist countries part of the European Union since 2004 (Bulgaria, Croatia, Czechia, Hungary, Poland, Slovakia, Slovenia, Romania). CEEc will be used as an abbreviation for Central and Eastern European cities and CEE for Central Eastern Europe.

ii   Excluding other CEEc.

iii   A classification of the Central Eastern European cities is proposed in this article based on population thresholds already used in many works (Bretagnolle et al., 2012, Bretagnolle et al., 2016). On this basis, 29 large cities (greater than 250,000 inhabitants), 185 medium-sized cities (50,000 to 250,000 inhabitants) and 620 small cities (between 10,000 and 50,000 inhabitants) have been identified in 2011.

iv   The *ORBIS* database produced by the Bureau Van Dijk (BVD) is a global database on capital links between mother and daughter companies. The extraction of *ORBIS*, available for this article, has been made upon a sample of the original database were a selection of the 3 000 biggest parent companies has already been applied, which reduces the number of possible subsidiaries. The uniqueness of *ORBIS* is to make available at city level, data on both companies controlling capital and on the companies they own, for the eight CEEc considered.

v   The turnover and participation in capital variables were used to quantify the foreign direct investment (FDI) income. We have then calculated an index of income intensity called FORCE (Śleszyński, 2007), which is proportional to the share of capital owned by the foreign firm (FORCE= $a\,b$, where $a$ is the rate of participation in the capital of a firm located in CEE by a foreign firm and $b$ is the income of the owned firm calculated on a turnover basis).

vi   In order to consider only the transnational links, this study does not take into account the links within a CEE country.

vii   For example, firms located in Ljubljana at the N-1 level are linked by a capital link towards firms in cities in N-2 in the European Union (Cyprus, Austria), CEE (Croatia and Czechia), but also from the former Yugoslavia (Montenegro, Serbia).

viii   Some other chains that are not connected to the main graph, generate a significant amount of investment income. This is the case of the links between Seoul (South Korea), Jászfényszaru (Hungary) and Galanta (Slovakia), relative to the Korean group Samsung Electronics, which was at the origin of the largest FDI revenues in CEE in 2013.

ix   The link between the cities of Bergamo (Italy), Koper (Slovenia) and Novi Sad (Serbia) is explained by the following capital control chain: Societa Italiana Acetilene e Derivati SPA (Bergamo) owns Instrabenz Plini D.O.O. (Koper), which itself controls the capital of Instrabenz Plini (Novi Sad). However, the latter company has also a subsidiary in Bakar in Croatia, which specializes in the distribution and production of industrial gases.

x   Łódź, with the highest betwenness centrality, is the headquarters of Pharmena S.A., a pharmaceutical and biotechnology company. It is controlled by General Electric Company (New York) and the Government of Norway (Oslo) – public actors, as well as Aviva PLC (London) and KBC Group SA (Brussels) – private actors.

xi   The sectors are the following*:* car industry *(*repair and sale of motor vehicles); finance/insurance/banking (life insurance, financial leasing); IT (IT services activities); industry (industrial production of chemical products, cement, textiles, plastics, household appliances, paper); media/advertising/communication (organization of the television program, radio, wireless telecommunications activities); real estate/tourism (real estate agency, hotels and similar housing); sales/trade (sale of machinery, chemicals, pharmaceuticals, textiles and food products); services/construction; energy.

xii   The calculation of all indicators was done under Tulip, software developed by the team of D. Auber and P. Mary of UMR LaBRI. The computation of the betwenness centrality is thus constrained by the way in which it realizes the software to know according to the algorithm of U. Brandes (Brandes, 2001).